\documentstyle[preprint,aps]{revtex}
\def\be {\begin{eqnarray}}
\def\ee {\end{eqnarray}}
\def\beq {\begin{equation}}
\def\eeq {\end{equation}}
\begin{document}

\draft
\preprint{NORDITA 95/47-N, {\tt nucl-th/9507023}}
\title{Kaon Energies in Dense Matter}
\author{V. R. Pandharipande$^{1}$,
C. J. Pethick $^{1,2,3}$ and Vesteinn Thorsson$^{2,}${\cite{byline}} \\}
\address{
$^{1}$ Department of Physics, University of Illinois at Urbana-Champaign,\\
1110 West Green St., Urbana, Illinois 61801-3080, USA\\
$^{2}$ NORDITA, Blegdamsvej 17, DK-2100 Copenhagen \O, Denmark \\
$^3$  Institute for Nuclear Theory, University of Washington,  Box 351550,
Seattle, Washington 98195-1550, USA}

\date{\today}
\maketitle
\begin{abstract}
We discuss the role of kaon-nucleon and nucleon-nucleon correlations in kaon
condensation in dense matter. Correlations raise the threshold density for kaon
condensation, possibly to densities higher than those encountered in stable
neutron stars.
\end{abstract}


\narrowtext

The possibility of kaon condensation in dense matter was suggested by Kaplan
and Nelson\cite{kn}, and it has been discussed in many recent
publications.\cite{bkrt,weise,tpl,blrt,leeetal,tw}.  The basic idea is that the
energy of a
$K^-$, $\epsilon_K$, is lowered by interaction with nucleons, and in neutron
star matter in beta equilibrium one expects negative kaons to be present if
the energy to create one in the medium is less than the electron chemical
potential, $\mu_e$, the energy required to add an extra electron.

One complication in determining the low energy interactions of kaons with
nucleons is that the interactions of kaons with protons {\it in vacuo} is
dominated by the $\Lambda(1405)$ resonance.  Fortunately this does not affect
the $K^-$ interaction with neutrons, which is expected to dominate in neutron
stars.  In dense matter the effects of the $\Lambda(1405)$ are likely
unimportant, because the kinematic region of interest is far from the
resonance.  To describe the $KN$ interactions Brown, Lee, Rho, and
Thorsson\cite{blrt}, herafter denoted by BLRT, employed an effective
Lagrangian based on chiral perturbation theory.  To circumvent the
difficulties with the resonance, parameters of the Lagrangian were fitted to
$K^+N, (N=n$ or $p)$, data, which are unaffected by the resonance.  The
Lagrangian was treated in mean field theory, and the density for condensation
was found to be $\sim 3-4 \rho_0$, where $\rho_0=0.16 {\rm fm}^{-3}$
is nuclear matter density.
This is to be compared with the central density of about
$4\rho_0$ for a neutron star of mass $1.4 M_\odot$ according to the estimates
of Wiringa, Fiks and Fabrocini{\cite{wff}} using realistic models of nuclear
forces.  The conclusion was that kaon condensation could affect the structure,
maximum mass and cooling rates of neutron stars significantly.

In this Letter, we first show that an important contribution to the kaon
energy in the BLRT treatment is the first term in an expansion in powers of
the particle density, with a strength proportional to kaon-nucleon scattering
lengths.  We then argue that kaon-nucleon correlations in the medium will be
reduced compared with those for a kaon and nucleon {\it in vacuo}, and that
consequently the attraction experienced by a kaon in matter will be smaller
that earlier estimates indicated.

BLRT use the Lagrangian of Jenkins and Manohar${\cite{jm}}$ containing terms
of the two lowest orders in the chiral expansion, and the nucleons are treated
in mean field theory, that is their densities are assumed to be uniform, and
equal to $\rho_n$ and $\rho_p$.  The spatially uniform state, which is the one
of lowest energy in this model, is given by
\beq
K^\pm = v_K e^{\pm i \epsilon_K t},
\eeq
and
\be
\nonumber
{\cal L} = (1 -\alpha)\dot K^+ \dot K^- -(m_K^2 -\delta^2) K^+K^-
\\-\frac{i}{2}V(K^-\dot K^+ - \dot K^- K^+),
\ee
with $V=(\rho_n+2\rho_p)/(2f^2)$, $\delta^2 =[\Sigma_{KN}(\rho_n + \rho_p)-C
(\rho_n - \rho_p)]/f^2$, and $\alpha =-[\tilde D(\rho_n + \rho_p) -\tilde
D'(\rho_n - \rho_p)]/f^2.$ The $V$ and $\delta^2$ terms describe interactions
of kaons with effective vector and scalar fields, while $\alpha$ takes into
account energy dependence of the interaction.  The pion decay constant, $f
(\approx 93 $ MeV), and the constant $C (\approx 33.5$ MeV) are well
determined, but the values of $\Sigma_{KN}, \tilde D,$ and $\tilde D'$ are
not well known.  The experimental $K^+N$ scattering lengths{\cite {bs}} are
used to determine $\tilde D$ for chosen values of $\Sigma_{KN}$ from Eq.(27)
of BLRT. Eq.(28) of BLRT gives
$\tilde D'=0.092/m_K$.  A large $\Sigma_{KN}$ implies a large energy
dependence for the $K^-N$ interaction, and the value $\Sigma_{KN}=278$
MeV was considered to be realistic by BLRT.  It is presently considered likely
that
$\Sigma_{KN}$ lies around 400 MeV \cite{leeetal}.  The corresponding values of
$\tilde D$ are $-0.23/m_K$ for $\Sigma_{KN} = 278$ MeV, and $-0.48/m_K$ for
$\Sigma_{KN} = 400$ MeV.

The $K^-$ energy may easily be calculated from Eq. (2), and is
\beq
\epsilon_K = \frac {1}{2(1-\alpha)}((4(1-\alpha)(m_K^2-\delta^2) +V^2)^{1/2}
-V),
\label{kaonenergy}
\eeq
as used by BLRT. Observing that $\alpha, \delta^2$ and $V$ are linear in the
nucleon densities, $\rho_N$, we expand $\epsilon_K$ in powers of $\rho_N$. To
second order in the densities one finds
\be
\nonumber
\epsilon_K=m_K-(\frac{V}{2} + \frac{\delta^2}{2m_K} -\frac{\alpha m_K}{2})\\
+\frac{V^2}{8m_K}-\frac{\delta^4}{8m_K^3}+\frac{3}{8}\alpha^2 m_K -
\frac{V\alpha}{2}-\frac{\alpha \delta^2}{4m_K}.
\label{secondorder}
\ee
The first term is just the kaon rest mass.  The second term may be expressed
simply by making use of the fact that the $K^-N$ scattering
lengths\cite{scattering} are given by
\beq
a_{K^-n}= \frac{-m_R}{4 \pi f^2 m_K}(\frac{m_K}{2} +\Sigma_{KN} -C +(\tilde D
-\tilde D')m_K^2),
\eeq
\beq
a_{K^-p}= \frac{-m_R}{4 \pi f^2 m_K}(m_K +\Sigma_{KN} +C +(\tilde D +\tilde
D')m_K^2),
\eeq
where $m_R=m_Nm_K/(m_N+m_K)$ is the reduced mass.  This term is thus
the Lenz potential,
\beq
\Delta\epsilon_{K, Lenz} = \frac{2\pi}{m_R}(\rho_n a_{K^-n} + \rho_p a_{K^-p}),
\eeq
the well-known result for the energy shift to first order in the density ( see
also Ref.\cite{tw}).  Thus for given values of the scattering lengths,  the
Lenz potential is independent of the choice of $\Sigma_{KN}$.  However, terms
of order $\rho_N^2$ and higher depend on  $\Sigma_{KN}$.  When  $\Sigma_{KN}$
is small, $\delta^2$ and $\alpha$ are small (see Table 1) and the higher order
terms  dominated by $V^2/(8m_K)$ are repulsive.  In contrast, when
$\Sigma_{KN}$ is large, $\delta^2$ is large, $1-\alpha$ can become zero at
relatively small densities, $\sim 10\rho_0$, and the higher order terms are
attractive.

The dependence of the kaon energy on nucleon density is illustrated in Fig.1
for a proton fraction of 0.1.  The curves labelled by $\Sigma_{KN}$ values are
obtained from Eq.(\ref{kaonenergy}).
For comparison, the Lenz approximation, which contains only the first two
terms in Eq.(\ref{secondorder}), is also shown.  The Lenz term gives a major
contribution, while other terms representing the energy dependence of
the interaction and relativistic effects may be significant.

The Lenz term is exact only at low density, and there are corrections to it
when it is no longer a good approximation to assume that the $K^-N$ relative
wave function, $\psi_{K^-N}$, is the same as for a pair of particles {\it in
vacuo}.  For s-wave scattering, $\psi$ at low energy is given by
\beq
\psi_{K^-N}(r)= 1-a_{K^-N}/r ~{\rm for}~ r> R,
\eeq
where $R$ is the range of the $K^-N$ interaction.  The corresponding
unperturbed wave function is simply unity.

To illustrate the effects, we represent the $K^-N$ interaction by potentials,
corresponding to the two possible isospin values, $I=0,1$.
The scattering lengths are given by
\beq
a_{K^-N} = \frac{m_R}{2\pi}\int d^3r {\cal V}_{K^-N}(r)\psi_{K^-N}(r),
\eeq
where ${\cal V}_{K^-N}(r)$ represents the $K^-N$ potential.  Since
$\psi_{K^-N}(r\!\!<\!\!R) > 1-a_{K^-N}/R$ when ${\cal V}_{K^-N}(r)$ does not
have a repulsive core, $2\pi a_{K^-N}/m_R$ is larger than the volume integral
of ${\cal V}_{K^-N}(r)$ by a factor larger than $1-a_{K^-N}/R$. The potential
${\cal V}_{K^-N}(r)$ is believed to be short ranged, because the vector and
scalar mesons that contribute to it have masses larger than the inverse of
hadron radii.  Hence ${\cal V}_{K^-N}$ has a range $R<1$ fm, and the
factor $1-a_{K^-N}/R$ is larger than 1.41 and 1.85 for $N=n,p$ with the
scattering lengths $a_{K^-n} =-0.41$ fm and $a_{K^-p} =-0.85$ fm given
by  Eqs.(5) and (6), and the parameters taken from BLRT.  These factors
represent the correlation between low-energy $K^-$  and nucleons {\it in
vacuo}, and when it is significantly different from unity, the Lenz term is
valid only at densities low enough that the radius, $r_0 $, of a sphere with a
volume equal to the average volume per particle, $1/\rho = 4 \pi r_0^3/3$ is
larger than both $R$ and $|a|$.

In order to make quantitative estimates, we fitted square well and Yukawa
potential to the scattering lengths.  For a square well of radius 0.7 fm, the
depths are 126.2 MeV for neutrons and 181 MeV for protons, while for R=1.0 fm,
the corresponding values are 49.5 and 75.6 MeV. For a Yukawa potential $V_0
\exp(-m_xr)/(m_x r)$, one finds $V_0 = -795$ MeV for neutrons and -1075 MeV
for protons if the mass, $m_x$, of the exchanged particle is taken to be that
of the $\rho$ meson, 770 MeV. In neutron stars, it is the $K^-n$ interaction
that is the more important, since the proton fraction of matter is small, and
in Fig. 2 we show $\psi_{K^-n}$ for the three potentials.  These exhibit the
enhancement of the wave function for small separations.

At higher densities, the kaon energy in matter should be calculated
either by summation of diagrams, as in approaches that evaluate the $K^-N$
interaction $G$ matrix and higher-order correlations, or from variational wave
functions of the type
\beq
\Psi =[\Pi_{i=1,A} f_{K^-N}(\vec r_{K^-} -\vec r_i)] \Psi_N(\vec r_1, ...\vec
r_A),
\eeq
where $f_{K^-N}$ represents the $K^- -$ nucleon correlation and $\Psi_N$is the
ground state of a system with $A$ nucleons.  At the large densities of
interest in the cores of neutron stars, $r_0<1$ fm.  When $r_0 <R$ , the kaon
is weakly correlated with nucleons, and $f_{K^-N}$ is unity to a good
approximation.  In this limit the Lenz potential is replaced by the Hartree
potential, and the $K^-$ energy is given by
\beq
\epsilon_{Hartree}= m_K +\int{[\rho_n {\cal V}_{K^-n}(r)
+\rho_p{\cal V}_{K^-p}(r)] } d^3r.
\eeq

The Hartree potential is less attractive than the Lenz one by a factor in
excess of $1-a_{K^-N}/R$.  In Fig. 1 we show estimates of the Hartree
contribution to the kaon energy for the potentials we fitted to the scattering
lengths, as described above.  The Hartree calculation does not include effects
of energy dependence of the interaction or of relativity, but since the
Hartree energy is closer to $m_K$ than is the Lenz term, we expect these
effects to be smaller than estimated from Eq.  (\ref{secondorder} ).

To investigate the validity of the Hartree approximation we now estimate second
order terms in the $K^-N$ interaction.  For simplicity, we neglect the effects
of the small number of protons.  In this case, the main coupling of the kaon is
to  neutron density fluctuations, since the kaon is spinless, and consequently
does not couple to spin fluctuations. The  contribution to the energy of a kaon
at rest is then given by
\beq
\Delta\epsilon_K^{(2)} =-\sum_{\vec q, l} |{\cal V}(q)|^2 \frac{|<l|\rho_{\vec
q}|0>|^2}{\epsilon_K(q) -\epsilon_K(0) +E_l -E_0}.
\eeq
Here $<l|\rho_{\vec q}|0>$ is the matrix element of the operator for the
Fourier transform of the neutron density between
the ground state, $0$, and an excited state, $l$, and $E_0$ and $E_l$ are the
energies of the states. The energy of a kaon with momentum $\vec p$ is denoted
by $\epsilon_K(p)$.  If one neglects the recoil energy of the kaon, one finds
an upper bound  for the energy shift in second order:
\be
\nonumber
-\Delta\epsilon_K^{(2)} < \frac{1}{2}\sum_{\vec q} |{\cal V}(q)|^2
\chi(q,\omega=0)\\
=\frac{1}{2} \int d^3r d^3r'{\cal V}(r) \chi(\vec r -\vec r'){\cal V}(r')
\ee
where
\beq
\chi(q, 0) =\sum_l2\frac{|<l|\rho_{\vec q}|0>|^2}{E_l -E_0}.
\eeq
is the static density-density response function for neutrons, and $\chi(r)$ is
its Fourier transform.  The recoil energy of the kaon is comparable in
magnitude to the excitation energy of the neutron liquid for the wavevectors of
importance, so the actual second order contribution to the energy is expected
to be considerably smaller than the bound. The long-wavelength static
density-density response function for neutron matter may be estimated by using
the thermodynamic identity $\chi(q \rightarrow 0, \omega
=0) = d\rho_n/d\mu_n$, together with many-body calculations of the neutron
chemical potential as a function of density.  The dependence of $\chi$ on
wavenumber is not well known, and to obtain an estimate of the second-order
term we shall replace $\chi(q)$ by its long-wavelength value.  At a density of
4$\rho_0$, the estimates of the second order term obtained by using values of
$d\rho_n/d\mu_n$ from Ref.\cite{fp} are 7 MeV (square well, $R=1$ fm), 16 MeV
(square well, $R=
0.7$ fm), and 46 MeV (Yukawa), and for higher densities the estimates are
smaller.  In all cases, these energies are considerably less than the Hartree
potential.

The second order contribution to the energy is reduced greatly due to the
strong repulsive interactions between nucleons.  Had we neglected
neutron-neutron interactions, the estimates of the second order term, which
would then reflect Pauli blocking of intermediate states in the repeated
scattering of a kaon by a neutron{\cite {lbmr}}, would have been much
greater.
For a free Fermi gas, $\rho_nd\mu_n/d\rho_n =2 E_F/3$, where $E_F \approx 58
(\rho_n/\rho_0)^{2/3}$ MeV is the Fermi energy.  At a density of 4$\rho_0$,
$\rho_nd\mu_n/d\rho_n$ estimated from the calculations of Friedman and
Pandharipande${\cite{fp}}$ is about 5 times larger than the free gas value.
The fact that stable neutron stars with a mass of 1.4 $M_{\odot}$ exist is a
clear indication that the equation of state of neutron star matter is
considerably stiffer than that for a free Fermi gas.  While there are still
uncertainties about the properties of matter at high densities, this increased
stiffness is a rather general property shared by many models, as one can see
from, e.g., Ref.\cite{wff}.  As a first step towards a more complete many-body
treatment, we are calculating the kaon energy using the lowest order
constrained variational (LOCV) method \cite{pb}, and the results will be
reported elsewhere.

To explore consequences of our calculations for kaon condensation, we also
show in Fig.1 the electron chemical potential, $\mu_e$, calculated by Wiringa
et al.\cite{wff}.  Our Hartree estimates of kaon energies always lie above
$\mu_e$ for essentially all stable neutron star models.  However, for a
number of reasons we are prevented  from
definitively ruling out kaon condensation in neutron stars.
First, we have used simple approximate forms for the kaon-nucleon
interactions, and estimates should be made with better forms of the
interaction, which should also take into account the variation of the coupling
constant, $f$, with nuclear density\cite{blrt}.
Second, corrections to the Hartree result need to be investigated in
detail.  Third, as one can see from the figure, estimates of $\mu_e$ at high
densities are subject to considerable uncertainty:  models of nucleon
interactions that fit laboratory data equally well lead to very different
values of $\mu_e$ in neutron star matter.  All these points need to be looked
into.

The main conclusion of our calculations is that, in a dense medium such as
that in the interiors of neutron stars, correlations are very different from
those for a dilute gas.  This has the effect of reducing the attraction
experienced by a kaon in dense matter.  While a low-order chiral expansion may
be useful for describing low-energy scattering of kaons by nucleons, this
information alone can predict the kaon energy only at low densities.  When the
$K^-N$ scattering lengths are as large as indicated by BLRT, the Lenz
approximation overestimates the attraction felt by a $K^-$ in neutron star
matter at the density of neutron star cores by roughly a factor of 2.

This work was supported in part by NSF grants NSF PHY94-21309 and NSF
AST93-15133, and NASA grant NAGW-1583 . We are grateful to D. G. Ravenhall, W.
Weise, and R. B. Wiringa for helpful discussions.  One of us (CJP) is grateful
to the Institute for Nuclear Theory at the University of Seattle for
hospitality during the workshop there on chiral symmetry, when he received
partial support from the Department of Energy grant to the Institute.

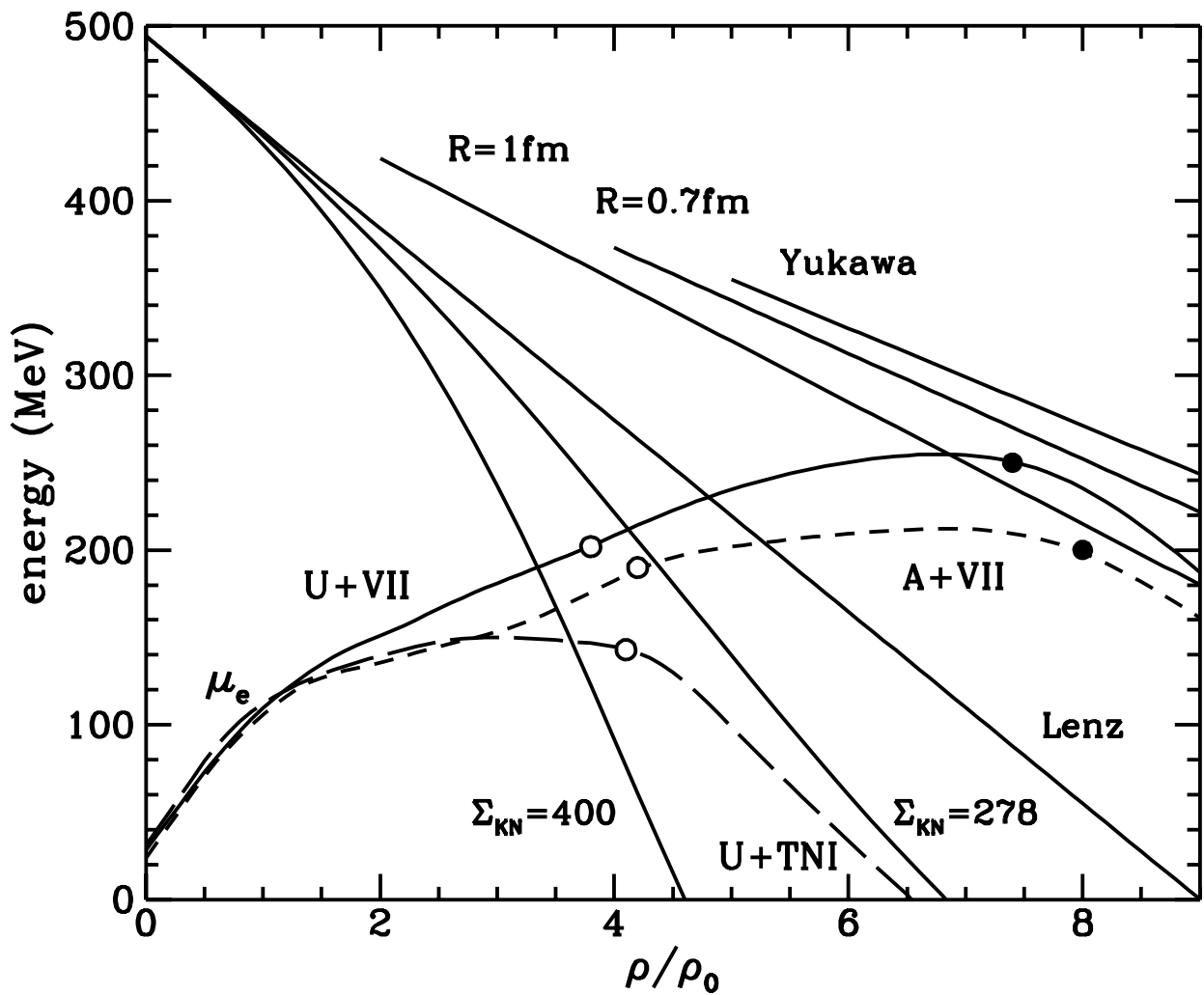
\begin{figure}

\caption{Energy of a single $K^-$ in matter as a function of density, and
electron chemical potentials taken from the calculations of Ref.[6].
The calculations of kaon energies are for a proton
fraction of 0.1. Curves labelled by values of $\Sigma_{KN}$, in MeV, are
evaluated from Eq.(3).  Hartree potentials, evaluated from Eq.(11),
are shown for square wells of radii $R=1$ fm and $R=0.7$ fm, and for a Yukawa
potential.  A and U
refer to the AV14 and UV14 two-body potentials, and VII and TNI to the UVII
and TNI many-body potentials.  The open circles indicate the central densities
of neutron stars with a mass of 1.4$M_{\odot}$, and filled circles indicate
the central density for the most massive stable neutron star for the given
equation of state. }

\label{energy}
\end{figure}

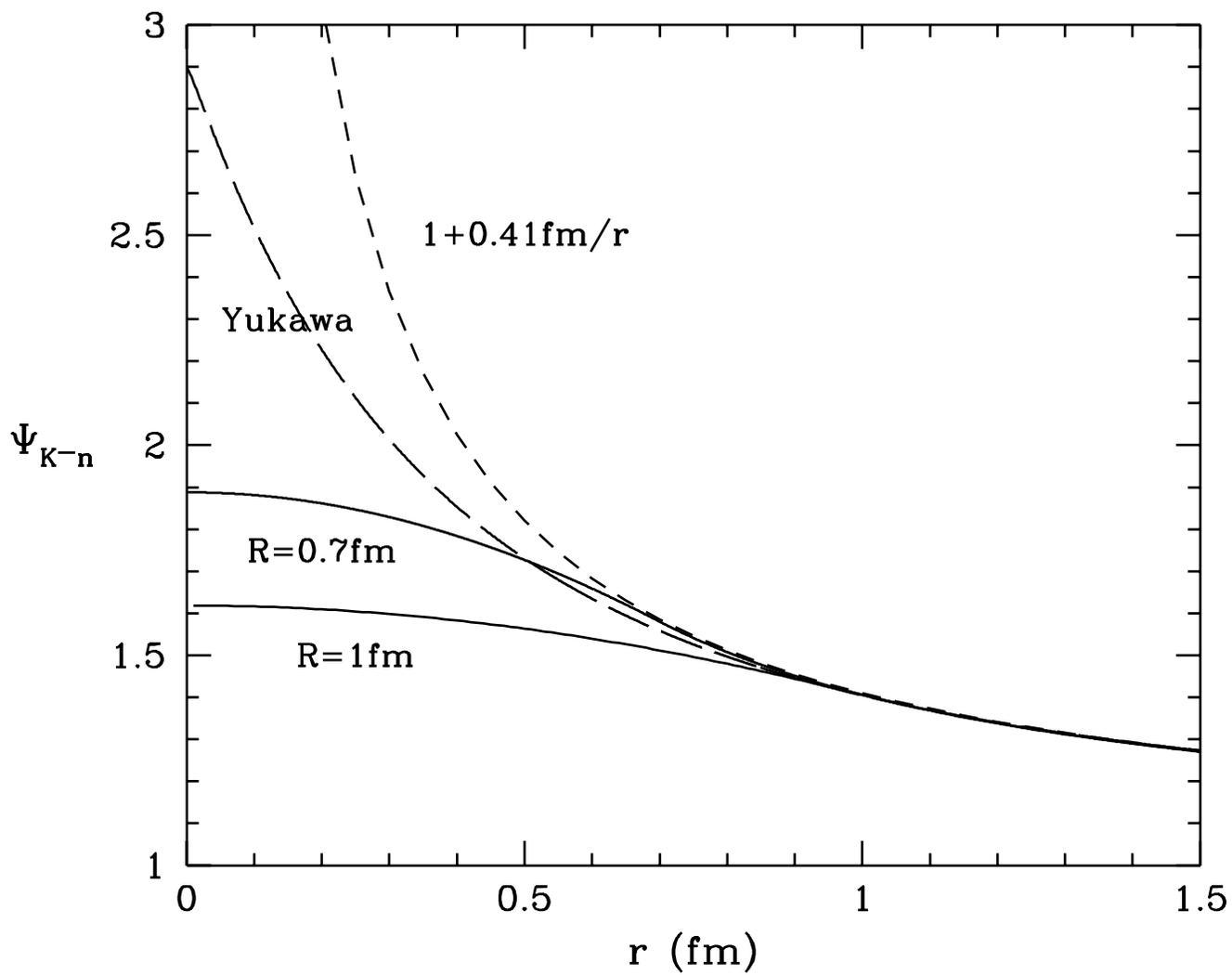
\begin{figure}

\caption{Wave function for the relative motion of a $K^-$ and a neutron at zero
energy for square well potentials (solid lines) and a Yukawa potential (
long-dashed line). The short-dashed line shows the asymptotic limit of
the wave functions.}
\label{wavefunction}

\end{figure}


\begin{references}
\bibitem{scattering} We caution the reader that, in order to make the paper
accessible to most physicists, we use the usual sign convention for scattering
lengths, with positive scattering lengths corresponding to repulsive
potentials.  This is the opposite of the convention commonly employed in the
literature on kaon-nucleon and pion-nucleon interactions.
\end{references}
\end{document}